%% file: Preprint.tex
\documentclass[article,,12pt,number,sort&compress]{elsarticle}

\usepackage{graphicx}
\usepackage{amssymb}
\usepackage{amsmath}
\usepackage{amsfonts}

\usepackage{lineno}
\usepackage{url}
\usepackage{hyperref}

\hypersetup{
		colorlinks=true,       		
    	linkcolor=blue,          	
    	citecolor=blue}

\usepackage[nameinlink]{cleveref}

\usepackage[labelfont=bf]{caption}
\usepackage{epsfig}
\begin{document}

\begin{frontmatter}


\title{Formation mechanism, stability and role of zinc and sulfur vacancies on the electronic properties and optical response of ZnS}



\author[1]{P.R.A de Oliveira}
\author[1]{L.Lima}
\author[1]{G.Felix}
\author[2]{P.Venezuela}
\author[1]{Fernando Stavale}

\address[1]{Centro Brasileiro de Pesquisas Físicas, 22290-180, Rio de Janeiro, RJ, Brazil}

\address[2]{Instituto de Física, Universidade Federal Fluminense, Campus da Praia Vermelha,
Niterói, RJ, 24210-346, Brazil}

\begin{abstract}
Combining experimental and theoretical tools, 
 we report that Zn vacancies play an important role in the electronic and optical responses of ZnS sphalerite. The defective surface of ZnS (001) single crystal prepared in ultra-high vacuum conditions, has been shown to exhibit a semiconducting character instead of the insulating properties of the pristine structure, as revealed by X-ray photoelectron spectroscopy (XPS). Interestingly, this effect is attributed to the formation of zinc vacancies in the ZnS system, which also alter the optical response of the material, as supported by photoluminescence (PL) measurements comparing pristine and S-rich (Zn-
poor) ZnS. To address these findings from a theoretical point of view, first-principles
calculations based on density functional theory (DFT) were performed. The optical
properties of cation-defective ZnS were evaluated using random-phase approximation
and hybrid functional DFT calculations. These calculations revealed absorption peaks in the visible range in the defective ZnS rather than solely in the ultra-violet range obtained for defect-free ZnS. The combination of this finding with joint density of states (JDOS) analysis explains the emergence of new luminescence peaks observed in the PL spectra of cation-defective ZnS. These findings highlight the role of Zn vacancies in tuning ZnS optical properties, making it a potential candidate for optoelectronic applications such as LEDs and photodetectors.
\vspace{0.25cm}
\end{abstract}


\end{frontmatter}



\input{Core_parts/Introduction}
\input{Core_parts/Methodology}
\input{Results/Experimental/Experimental}
\input{Results/Theoretical/Theoretical}

\input{conclusion}

\input{FurtherParts/AuthorContribution}
\input{FurtherParts/AuthorDeclaration}
\input{FurtherParts/Acknowledgments}
\input{FurtherParts/SupportingInformation}
\input{FurtherParts/DataAvailability}

\bibliography{Reference.bib}






\end{document}

%% file: Core_parts/Introduction.tex
\section{\label{sec:Introduction}Introduction}

Zinc sulfide (ZnS) is an II-VI group semiconductor usually arranged in two crystallographic configurations: zinc blende (zb) and wurtzite (wz)
\cite{shahi2018accurate,grunwald2012transferable,xiong2007tunable,wright2004interatomic}. In addition, other structures such as rock-salt \cite{ulian2019thermomechanical}, free-standing, and graphene-like 2D-ZnS \cite{tse2021electronic,lashgari2016electronic} are theoretically predicted. The scientific interest in ZnS has risen in recent years due to its
role as a potential candidate for applications in photodetectors and optoelectronic devices
\cite{wang2012gas,geiger1913lxi}. In addition, ZnS is considered a promising material for photocatalysis,
particularly in processes related to $CO_{2}$ photoreduction and $H_2$ trapping \cite{zhu2022recent,Mamiyev2022,Meng2022}. However,
its wide band gap ($3.6$ eV) and high resistivity (in the range of \textit{petaOhms} ( $p\Omega$ )) imposes
several challenges for its use in large scale industrial applications \cite{Meng2022}

Defect engineering is a common strategy to overcome these challenges \cite{WoodsRobinson2020,Bai2018,Yan2022}.
This mechanism tunes electronic and optical properties through new doping states arising
inside the wide ZnS band gap \cite{Yan2022}. As discussed in recent investigations, either intrinsic
defects, such as zinc vacancies ($V_{Zn}$) and sulfur interstitial ( $S_{i}$ ), as substitutional
impurities like transition metal and rare earth elements, play a crucial role in broadening
optical absorption and enhancing molecular absorption efficiency \cite{Kurnia2015,Liu2020,Saleh2019}. Of particular
interest are cation vacancies generated by the removal of zinc atoms \cite{Yan2022}. This strategy can
facilitate the reduction of $CO_{2}$ molecules in wz-ZnS \cite{Luo2023} and enable light emission in the visible range \cite{hao2018zinc}, making cation-doped ZnS structures potential candidates for photocatalysis, photonic and light-emitting applications \cite{thompson2023red,shao2025strong,jiang2024high}. Despite these potential applications, experimentally controlling the amount of zinc vacancies, while simultaneously managing impurities such as carbon, nitrogen, or oxygen during doping, remains challenging and may compromise the experimental
analysis \cite{hao2018zinc}. This scenario might explain the lack of experimental realization coupled
with theoretical tools, to grasp the impact of cation defects on electronic and optical
properties of ZnS structures in the most stable phase (zb).

In this work, we investigated the role of zinc vacancies in the electronic and optical properties of zb-ZnS (001) single crystals. We combined ultra-high vacuum X-ray photoelectron spectroscopy (XPS) analysis and ex-situ photoluminescence (PL) experiments with first-principles calculations based on the density functional theory (DFT) framework. Our XPS experiments revealed that zb-ZnS exhibits enhanced conductivity as the zinc content decreases relative to sulfur. The suppression of charging, leading to the XPS peaks appearing at the expected binding energy positions, highlights this effect. Notably, a shift toward lower binding energy of the valence band maximum is observed in the cation-doped system compared to the defect-free sample. This shift is attributed to the formation of a stable p-type structure induced by zinc vacancies. Interestingly, a green-like emission is observed when the sample is irradiated with an external laser source ($2$ mW, $405$ nm), indicating that zinc vacancies in ZnS might also enhance the optical properties of the system. Additionally, PL
experiments reveal a new emission peak in the visible range centered at $588$ nm for the
cation-doped system, confirming the enhanced optical response upon doping. This
experimental peak can be deconvoluted into two components at $584.3$ nm and $605.0$ nm. These
components arise due to the recombination process of electrons and holes
within defective levels provided by the removal of zinc atoms. Our DFT calculations
support the experimental findings. Firstly, our  calculation of some
native defects in ZnS reveal the Zn vacancy system as the most energetically favorable one. In the electronic structure point of view, the calculations show the formation of zinc vacancies gives rise to a p-type system, characterized by transition energies that occur when the electronic chemical potential is close to the valence band. Moreover, this
configuration introduces shallow levels into the gap and displays a band-gap narrowing, which might explain the better
conductive behavior of cation-doped samples in the experiments. On the other hand, the
optical properties of ZnS are discussed in the light of the random phase approximation
(RPA) \cite{onida2002electronic}, employing hybrid functional DFT calculations to avoid
underestimation of the calculated absorption spectra and joint density of states (JDOS)
\cite{Becke1996}. In this scenario, cation-doped ZnS structures display a broad range of
absorption peaks and emission lines. In particular, charged $V_{Zn}$ allows for new absorption
peaks in the visible range and two new emission lines, also in the visible range, in good
agreement with the experimental results.

%% file: Core_parts/Methodology.tex
\section{\label{sec:Methods}Methodology}
\subsection{Experimental methods}

The ZnS single crystals (purity $99.999$$\%$ )purchased from Maltech gbmh were prepared in situ
via several cycles of argon sputtering ( $E = 600$ eV, $I_S = 5\mu A$, $t= 5$ min) and annealing at different
temperatures (up to $1520$ K, during $30 $ min). X-ray
photoelectron spectra (XPS) were measured in a Near Ambient Pressure (NAP) system equipped with a \textit{SPECS PHOIBOS} NAP 150 hemispherical analyzer and a multiple differential pumping stages that allow for ultra-high vacuum (UHV) conditions in the samples analysis chamber to be kept at base pressures better than ($1\times 10^{-9}$ mbar). The XPS spectra were acquired using a monochromatic $Al-k\ \alpha $ ($1486.6$ eV) and the photoemission spectra were collected from a sample surface area $ < 1 \ mm$. All the XPS spectra were measured at normal emission (i.e the sample surface plane is
perpendicular to the analyzer acceptance angle). The binding energy was calibrated in
reference to the $Ag \ 3d_{5/2}$, which was set at $368.2$ eV for an Ag (001) single crystal. A pass
energy of $50$ eV and $30$ eV for the survey and high-resolution spectra, respectively, was set in the
experiments. All the spectra were analyzed through the Casa XPS software \cite{Fairley2021}. The
Tougaard background was used with a Gaussian-Lorentzian GL ($30$) mix to fit the
components from the high-resolution spectra. All XPS spectra are shown without any binding energy calibration. 

The optical properties of the ZnS single crystal were carried out using an ex-situ home-
built Photoluminescence (PL) spectroscopy setup. The PL spectra were obtained at room
temperature in backscattering geometry using a home-built system equipped with an
Andor Shamrock spectrometer within idUS charge-coupled device (CCD) detector. The
measurements were obtained using a $488$ nm laser with a spot diameter better than $1 \ \mu m$ and
power of $400 \ \mu m$.The spectrometer spectral broadening for this configuration was
determined using a silicon wafer peak at $520 \ cm^{-1}$ fitted using a Gaussian line shape with a full-width-half maximum (FWHM) of $6 \ cm^{-1}$. The PL data were fitted by a Voigt
function, mixing Gaussian and Lorentzian line shape.

\subsection{Computational methods}

 The ZnS system with and without defects were theoretically investigated through first-principles calculations based on the density functional theory (DFT), using the Quantum Espresso software \cite{Giannozzi2009}. For geometric optimization and electronic structures insights, the
calculations were performed within the generalized gradient approximation (GGA), following the description Perdew, Burke and Enzenhorf \cite{Perdew1996} and employing ultra-soft pseudo-potential \cite{Vanderbilt1990} to solve the Kohn-Sham equations. The equilibrium
lattice parameters were obtained through calculations with a $6 \times 6 \times 3$ \textbf{k}-mesh for the Brillouin zone integration using a Gaussian smearing function of $\sigma= 0.05$ eV. For modeling the system with and without defect a ($2 \times2 \times  1$) supercell was constructed. The defective ZnS structures were obtained by removing one atom
(vacancy defect) from the supercell, thus keeping the doping limit lower than $5\%$. The kinetic energy of the wave functions and density charge cutoffs were $38$ Ry and $456$ Ry, respectively. The calculations were considered converged when the Hellman-Feynman forces on each atom were less than $0.0001$ eV $/$ \AA

To improve the description of the optical properties calculations, a mixing with Hartree and Fock to describe the exchange-correlation energy based on the $PBE-\alpha$ hybrid functional theory \cite{Becke1996} was applied. In this regard, Norm-Conserving pseudo- potentials were used \cite{Hamann2013}. In our study, we employed $\alpha=0.27$, which was the best value matching the experimental band gap of $3.6$ eV. Besides the energy cutoff of the wave functions kinetic energy in the reciprocal space, a cutoff on the Fock space with the same value was added to accelerate the convergence of the calculations. The optical properties were evaluated within the dipole approximation, following the random-phase
approximation \cite{onida2002electronic}. The absorption spectra are obtained by computing the average of the imaginary
part of the dielectric tensor, which is obtained according to the Kramers-Kronig relations \cite{Karazhanov2007} (see S4). The JDOS was calculated employing a Gaussian occupation with the same smearing $\sigma$ to ensure methodological consistency.

%% file: Results/Experimental/Experimental.tex
\section{Experimental Results}
\subsection{XPS analysis}

\begin{figure}[h!]
    \centering
\includegraphics[width=1.0\textwidth]{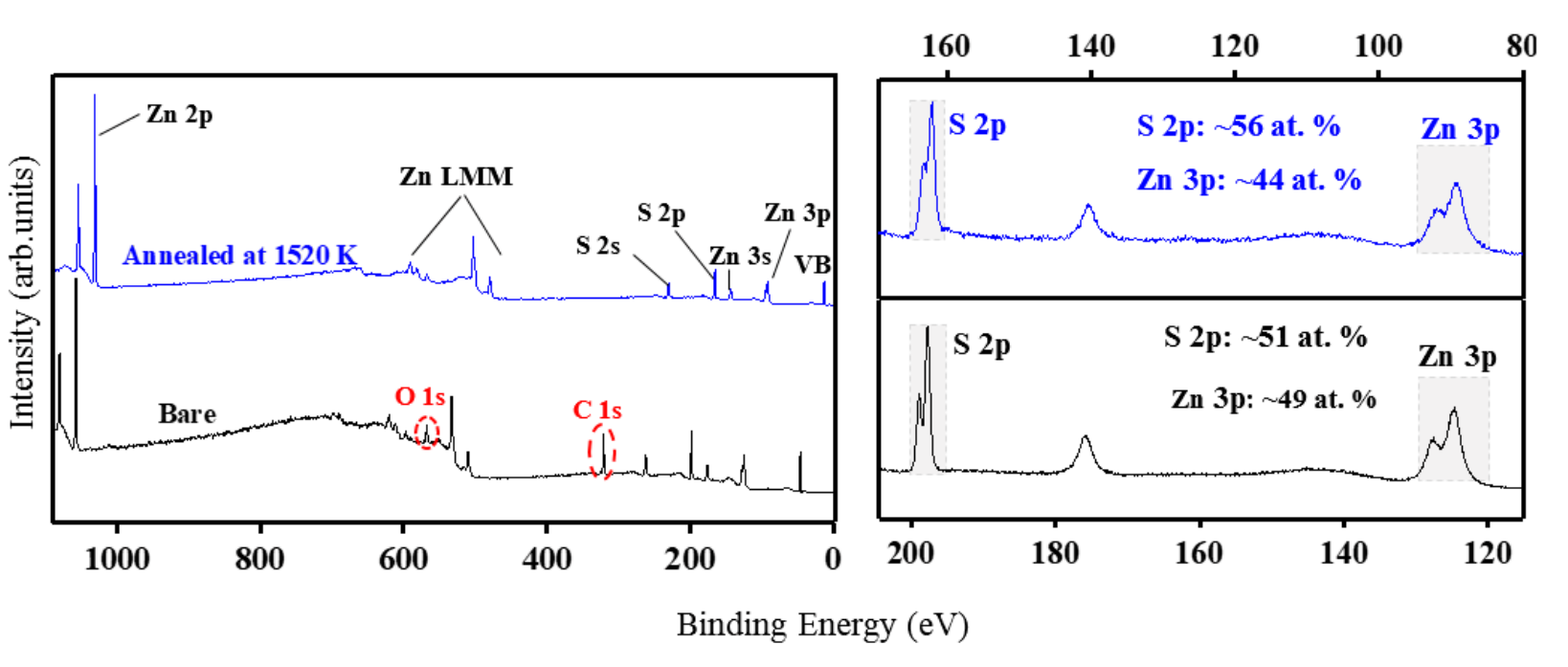}
    \caption{XPS spectra of bare and UHV annealed ZnS single crystal. (a) survey scan and (b) small region analysis spanning from S 2p to Zn 3p. All the spectra were acquired at room temperature and at normal emission.
}
    \label{fig1}
\end{figure}

Fig.\ref{fig1} shows the XPS spectra of the ZnS single crystal before and after in-situ surface preparation. In Fig.\ref{fig1}(a), the effect of the annealing temperature on the binding energy position of the ZnS main components is highlighted. In the bare sample scenario, that is,the sample measured as loaded, the XPS components appeared almost $35$ eV offset from the expected positions. For instance, Zn 2p, C 1s, and S 2p are identified at $1058$ eV, $320.8$ eV, and $197.8$ eV, while in the literature accepted values are $1022$ eV, $284.8$ eV and $161.8$ eV, respectively \cite{Dengo2020}. These binding energy shifts are explained by the insulating nature of pristine ZnS systems \cite{Barreca2002,Oliveira2024}. The absence of an effective electron-hole recombination rate results in strong electrostatic charging in the sample, thus reducing the kinetic energy of the emitted photoelectrons. One effective strategy to mitigate this scenario is in-situ preparation at high annealing temperatures \cite{Oliveira2024}. Indeed, several annealing cycles at $1520$ K allowed for a shift toward lower binding energies, as shown in the blue survey spectrum in Fig.\ref{fig1}(a). Here, Zn 2p and S 2p are located at $1022.8$ eV and $161.8$ eV, in good agreement with the binding energies of Zn atoms bound to S species in the $Zn^{2+}$-$S^{2-}$ configuration. Contamination species like carbon or oxygen were not found because UHV annealing removes such initially adsorbed surface species. Additionally, as we recently reported, this procedure might induce defect formation in the system, increasing the number of majority carriers and suppressing the sample charging. In that case, the ZnS surface displayed a more conductive character, and the XPS components were identified at the expected binding energies\cite{Oliveira2024}. To gain insights into the nature of the defects (cationic or anionic), the relative atomic concentration between S 2p and Zn 3p in the bare and UHV-annealed samples are shown in Fig.\ref{fig1}(b). It is worth noting that the S 2p and Zn 3p photoelectrons are ejected from equivalent depths, and their relative sensitivity factors, following the \textit{Scofield} table, are similar \cite{Scofield1976}. Therefore, the relative atomic concentrations obtained from these peaks are helpful in following the surface modifications occurring due to the sample UHV treatment. As discussed, prior in-situ treatment the binding energy shift of the main components is close to 35 eV. While for the bare sample the S 2p and Zn 3p components are found at $196$ eV and $125$ eV, respectively, these components after the UHV annealing at $1520$ K are located at $161.8$ eV and $90$ eV. Although there is a large binding shift, it does not play a role in atomic species quantification. In this case, we can evaluate the relative atomic concentration of sulfur and zinc in each scenario without losing generality. While in the pristine scenario (insulating sample) an almost $1:1$ ratio is found, the UHV annealed sample reveals an increase in the sulfur content by up to $5$ at.\% relative to zinc atoms. This finding suggests a sulfur-rich surface, which is attributed to the formation of a cation-defective system.

\subsection{Electronic structure and Optical response analysis}

\begin{figure}[h!]
    \centering
    \includegraphics[width=1.0\textwidth]{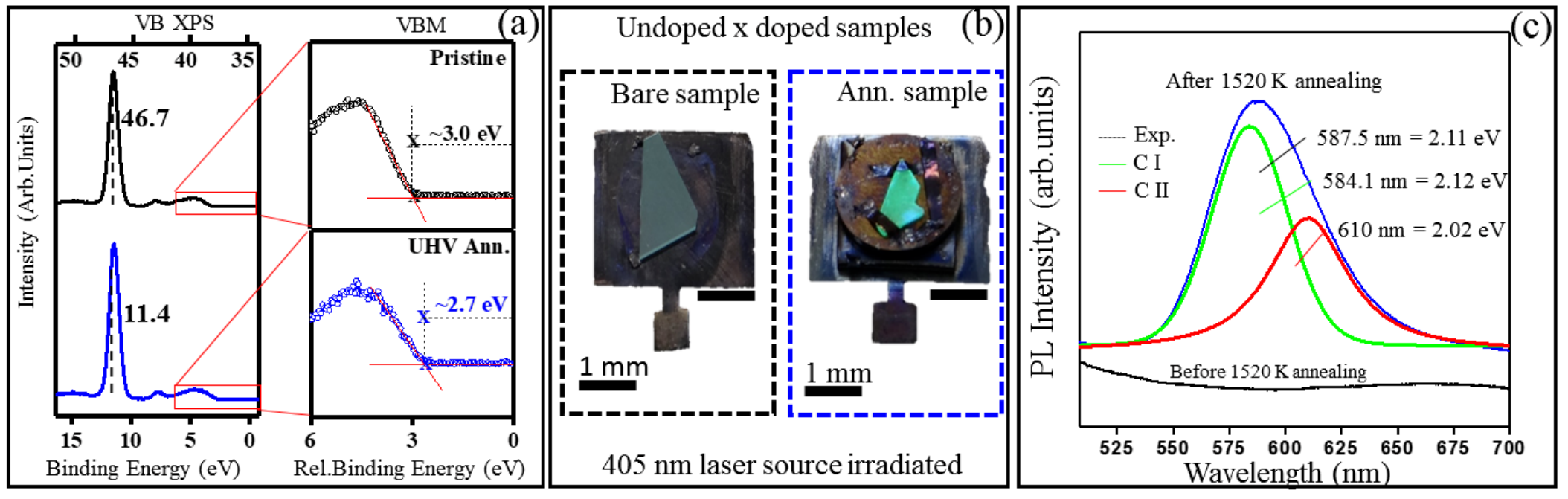}
    \caption{(a) XPS Valence band spectra of bare (left panel) and UHV annealed (right panel) ZnS single crystal; (b) Undoped and doped ZnS single crystal appearance upon laser irradiation. (c)Ex-situ PL spectrum before and after UHV annealing at 1520 K. The XPS spectra are shown as loaded. All the spectra were collected at room temperature
}
    \label{fig2}
\end{figure}
More information regarding the role of cation defects in the electronic and optical responses of zinc blende ZnS is shown in Fig.\ref{fig2}. In Fig.\ref{fig2}(a) (left panel), the valence band spectra of bare and in-situ prepared at 1520 K ZnS are depicted. It should be noted the Zn 3d peak in the bare sample is significantly shifted. The typical binding energy in ZnS sample is close to $11.4$ eV, while in this scenario it rises at $46.7$ eV, configuring a B.E shift of $35.2$ eV, compared to the expected value. The UHV annealed sample, on the other hand, does not reveal any binding energy shift, as depicted in the bottom of Fig.\ref{fig2}(a) (left panel). This sample seems to be more conductive due to the removal of zinc atoms, as discussed later. In this scenario, the main component related to Zn 3d orbital is found at
$11.4$ eV, as expected. More insights into the electronic structure were obtained comparing the valance band maximum (VBM) of both samples, as displayed in right panel of Fig.\ref{fig2}(a). These measurements are regarding to the Fermi-level, which means they are reliable even for the bare sample,
because the relative binding energy is considered. In this scenario, the VBM of the bare sample was found to be close to $3$ eV, while the UHV annealed sample VBM displayed a slight shift toward low binding energy, with the VBM arising close to $2.7$ eV, as noted on the top and bottom of the Fig.\ref{fig2}(b) (right panel), respectively. This finding is a general trend for p-type semiconductors. In ZnS, the most likely way to obtain a p-type system is through the formation of either zinc vacancies or sulfur interstitials. However, only the former one would give rise to an S-rich surface, which seems to be the ZnS scenario after UHV treatment as discussed in Fig. \ref{fig1}(b). An interesting strategy to probe the nature of the semiconductor (n or p) is the laser-assisted XPS measurements, as recently reported\cite{Baer2020}. Indeed, UV-laser assisting XPS measurements is a well-known method to mitigate sample charging \cite{Copuroglu2013}. Due to the low wavelength, it can excite electrons near to the surface, which enhances the electron-hole recombination rate, hence mitigating the positive electrostatic field generated due to the photoelectron emission. This aspect is highlighted in Fig S2-(a), where the binding energy of the bare sample is almost corrected due to the incidence of a $405$ nm laser source \cite{Copuroglu2013}. However, in the presence of doping (either native defects or substitutional impurities), two additional phenomena can be observed in laser-assisted XPS measurements: under-compensation (left-hand binding energy shift) and overcompensation (right-hand binding energy offset).

The direction is dependent on the nature of defects, with n-type and p-type standing for higher and lower binding energy shifts, respectively \cite{Baer2020}. In our case, the VB band measurement of the UHV annealed sample is slightly shifted toward low binding energy (see Fig.S2-(b) in the SI), indicating the p-type nature of the system.
However, a detailed investigation on this subject must be carefully considered because the total offset seems to be dependent on both wavelength and laser power (see SI),  though it is beyond the scope of this work.
Nonetheless, a quite interesting aspect to note – which is independent of those aforementioned parameters - is the ZnS color appearance under illumination of the $405$ nm laser. If the bare ZnS does not reveal any color-like appearance (Fig \ref{fig2}(b) left panel), the annealed sample, on the other hand, displays an intense green color aspect (Fig.\ref{fig2}(b) right panel). In view of the wide band gap of ZnS (at about $3.6$ eV), the visible emission observed in UHV annealed sample under a $405$ nm ($3.06$ eV) irradiation might signal an optical response enhancement due to the formation of cation defects. This aspect was quantitatively explored by photoluminescence measurement performed ex-situ. To this end, the ZnS samples were transferred from the XPS UHV chamber to our PL experimental setup by exposing them to ambient conditions followed by immediate PL measurements. Fig. \ref{fig2}(c) shows the PL spectra obtained for the bare, and UHV annealed ZnS samples. Noticeably, the bare sample does not display any photoluminescence response in the visible range, as indicated by the black spectrum. In general, ZnS can exhibit a near band-edge emission in the UV range (around $3.6$ eV) when excited with the appropriate light source, and its optical response could reportedly be modified by including transition and rare earth metal dopants \cite{Hu2006,Hoang2022}. In our PL experimental setup, however, we can only excite the sample with $488$ nm ($2.54$ eV), and investigation of the sample near band-edge emission is beyond the scope of this work. Yet, the PL response using our setup is helpful for identifying the sample’s optical properties modifications promoted by the UHV annealing, as indicated by the blue spectrum.

In Fig.\ref{fig2}(c), the UHV annealing results in an intense and broad peak located at approximately $587.5$ nm (spectrum in blue). The PL peak analysis can be performed employing two components, labeled as C1 (green line) and C2 (red line) located at $584$ nm and $610$ nm, respectively. Note worth, in our previous study on the ZnS (001) surface it was identified at similar UHV annealing conditions the formation of a stable zinc-terminated surface in which relative changes in the S/Zn atomic concentration were linked to the improvement of the surface conductivity due to the potential formation of Zn vacancies. However, the observed changes in the ZnS sample color appearance (Fig.\ref{fig2}(b) right panel) and the development of the photoluminescence response shown in Fig.\ref{fig2}(c) suggest that the formation of the Zn vacancies may indeed extend to the ZnS bulk counterpart and be connected to the PL response. In general, cation defects such as zinc vacancies might yield few states just above the VBM. These states can behave like shallow acceptors, trapping electrons and promoting the recombination process, which gives rise to new emission lines via radioactive recombination \cite{Zott1997}. To rationalize these experimental findings, in the following paragraphs, we have explored through theoretical calculations the stability and electronic properties modifications related to the formation of Zn and S vacancies in the ZnS bulk system.

%% file: Results/Theoretical/Theoretical.tex
\section{Theoretical Results}
\subsection{Point defect stability and electronic structure}

\begin{figure}[h!]
    \centering
    \includegraphics[width=1.0\textwidth]{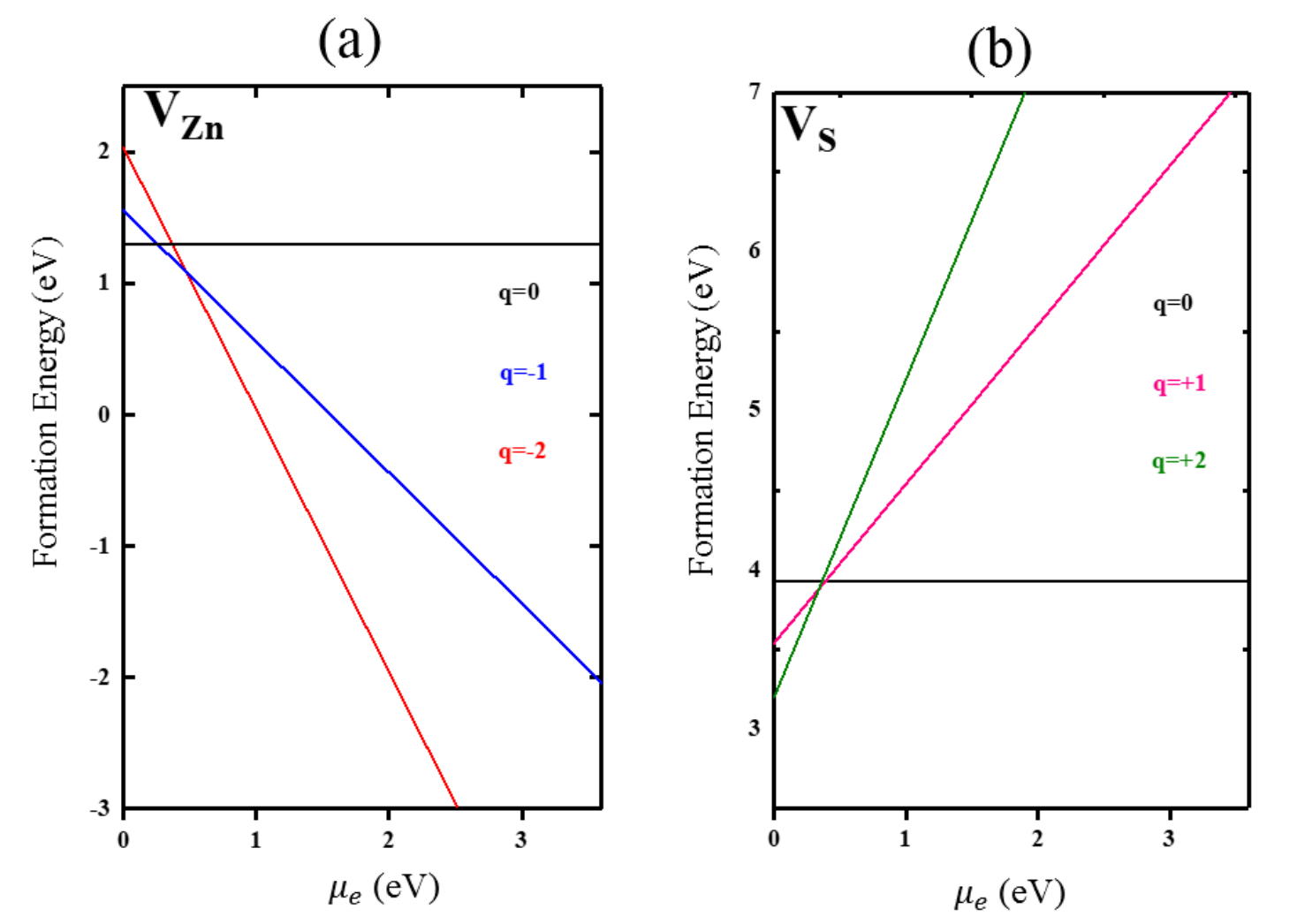}
    \caption{Formation Energy of (a) Zinc and (b) Sulfur vacancy as a function of the electronic chemical potential in an S-rich scenario.}
    \label{fig-Form}
\end{figure}

The formation energy of defect $D$ in a particular charge state $q$ ($E_{form}(D^{q})$) is a well-known strategy for describing the stability of point defects \cite{Janotti2007}. Considering the different charge states ($q= -2,-1, 0, 1, 2$) for zinc and sulfur vacancy in an S-rich scenario we computed the formation energy (See SI) as a function of the electronic chemical potential $\mu_{e}$ as depicted in Fig. \ref{fig-Form}. In the neutral defect picture (i.e, $q = 0$), zinc vacancy is more stable than sulfur vacancy ($E_{form}(V_{Zn}^{0}= 1.3 \ eV)$ , $E_{form}(V_{S}^{0}= 3.9 \ eV)$).  In addition, With $\mu_{e}$ pinned at the top of valence band, we obtained $E_{form}(V_{Zn}^{-1}) = 1.56 \  eV$, $E_{form}(V_{Zn}^{-2}) = 2.04 \ eV$, $E_{form}(V_{S}^{+1}) = 3.54 \ eV$,$E_{form}(V_{S}^{+2}) = 3.20 \ eV$, according to Fig.\Ref{fig3}(a) and Fig.\Ref{fig3}(b), respectively. These findings are well-aligned with the general trend  previously reported \cite{Yao2012}, which shows that charged zinc vacancies are more stable than charged sulfur vacancies. Furthermore, in this same scenario ($\mu_{e}$ pinned at the top of the valence band), the formation energy of charged zinc vacancy, especially the $E_{form}(V_{Zn}^{-1})$, is close to that of neutral zinc defects, which suggests that both nature (neutral and charged) of cation defect are equally stable on ZnS and might coexist. On the other hand, according to our calculations, it is expected that if sulfur vacancies are formed on ZnS, it should stay in the charge state $+2$. Overall, the cation-defect ZnS is the most energetic favorable configuration in our system

 An interesting parameter to investigate the nature of the stability (n-type or p-type) in the $V_{Zn}$-ZnS picture is the transitions $\epsilon(q/q')$ highlighted in Fig. \ref{fig-Form}(a)-(b) by the crossing of each line. This parameter computes the particular $E_{fermi}$ value where a transition from a defect in the charge $q$ to $q'$ takes place \cite{Janotti2007}. Zinc vacancy behaves as an acceptor, thus decreasing in energy as $\mu_e$ increases. According to the formation energy calculations, all the possible transition states -$\epsilon(-2/0)$, $\epsilon(-1/0)$, and $\epsilon(-2/-1)$- occur slightly above the VBM. This finding illustrates that zinc vacancies in an S-rich ZnS scenario behaves as shallow-acceptors and are indeed stable as a p-type configuration.
\subsection{Electronic Structure Analysis}

\begin{figure}[h!]
    \centering
    \includegraphics[width=1.0\textwidth]{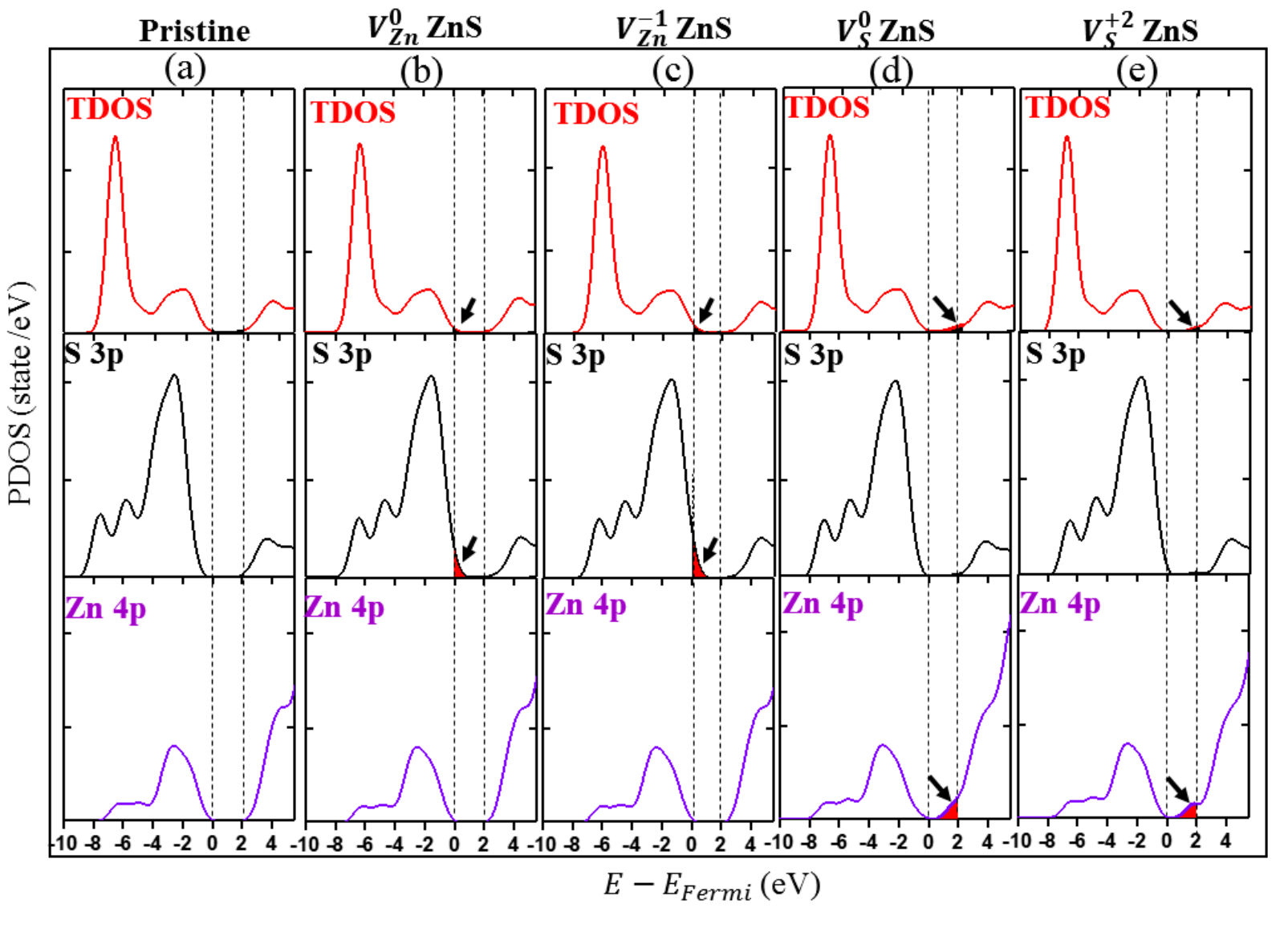}
    \caption{Projected Density of states of (a) pristine, (b) $V_{Zn}^0$,(c) $V_{Zn}^{-1}$, (d) $V_{S}^{0}$ and (e) $V_{S}^{+2}$ doped ZnS.The Fermi level was shifted to $0$ eV for clarity.}
    \label{fig3}
\end{figure}

Fig.\ref{fig3} shows the projected density of states of pristine ZnS and ZnS with vacancies. All the structures were calculated taking into account spin-polarization. The final magnetization is zero though. In this sense, Fig.\ref{fig3} shows spin-up projected density of states of such structures. The calculated band gap of the pristine ZnS is 2.2 eV, which is an underestimated value as expected from calculations through PBE-GGA methodology \cite{Perdew2017}. Regarding the total density of states of each system, we found similar characteristics. Firstly, the Zn 3d states dominate the portion far from the valence band. Conversely, the valence band is primarily composed of S 3p states, while the conduction band is mainly formed by Zn 4p and unfilled S 3p states. The main difference between defect-free and defective ZnS is the appearance of a few states inside the gap, indicated by the black arrow in Fig.\ref{fig3}(b)-(e). For instance, after the formation of zinc vacancy (Fig.\ref{fig3}(b)-(c)) there is a rise of a few states slightly above the Fermi-level. This finding is attributed to the decrease in the coordination number of S atoms due to the removal of zinc species, which promotes new S 3p-like states above the Fermi level, as highlighted by the filled red area in Fig.\ref{fig3}(b)-(c) (middle plot). In addition, the band gap is tailored to $~ 2 \ eV$. Although the aforementioned band gap underestimation through PBE-GGA calculations, this trend might indicate an electronic response improvement due to the formation of cation defects. The slight difference between neutral (Fig.\ref{fig3}(b)) and charged (Fig.\ref{fig3} (c)) cation-doped ZnS is the density of such S 3p-like states arising inside the forbidden zone, which seems higher in the case of charged zinc vacancy. Overall, both structures could explain the
enhancement of the electronic and optical response, as pointed out from XPS and PL experiments. On the other hand, the formation of sulfur vacancy gives rise to a few states slightly below the conduction band minimum, as revealed in Fig.\ref{fig3}(d)-(e). The decrease of the Zn coordination due to the removal of sulfur results in some Zn 4p-like states, highlighted by the filled red area in Fig.\ref{fig3}(d)-(e) (bottom panel). However, we did not observe a band gap reduction in either scenario (neutral or charged sulfur vacancy), indicating sulfur vacancies do not significantly improve the electronic response of the system.  This finding might be addressed to the doping concentration limit. In general, sulfur vacancies in zinc blende ZnS are predicted to play a role in electronic and optical properties only at high defect concentration (higher than $25 \%$) \cite{Liu2020}. As mentioned in the methodology, we maintained the doping below 5 $\%$, making sulfur vacancy unlikely to explain our experimental findings. Based on our theoretical calculations, only neutral and charged zinc vacancy align well with the experimental results, particularly regarding the enhancement of the electronic and optical response and the p-type nature of the system. Therefore, we next focus exclusively on these structures to evaluate the optical properties of doped ZnS from a theoretical perspective.


\subsection{Optical properties calculations}

\begin{figure}[h!]
    \centering
    \includegraphics[width=1.0\textwidth]{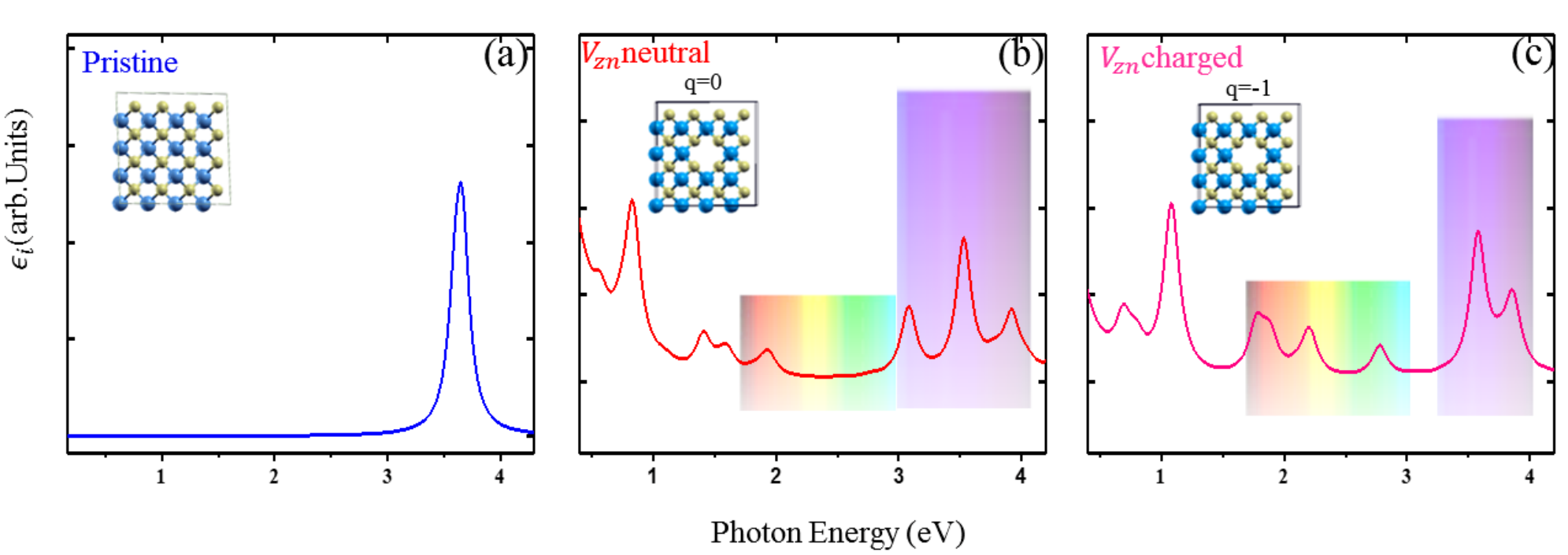}
    \caption{ Theoretical absorption spectra of of (a) pristine, (b) neutral $V_{Zn}$  and (c) charged $V_{Zn}$  ZnS
}
    \label{fig4}
\end{figure}

A broad optical absorption range is a key prerequisite for a material to be considered a potential candidate for photocatalysis and optoelectronic devices applications, as in the case of ZnS. In this sense, Fig.\ref{fig4} compares the optical absorption of ZnS in three particular scenarios: pristine, neutral p-doped, and charged p-doped ZnS. In the first case, the only observed peak was the absorption edge at $3.6$ eV, as shown in Fig.\ref{fig4}(a). This peak corresponds to incident photons with energies matching the ZnS band gap. This result is in good agreement with the experimental band gap of ZnS, as expected, given that all the optical property calculations were performed using $PBE \ -\alpha $  hybrid-functional, with  $\alpha = 0.27$ chosen to fit the experimental band gap value. Conversely, due to the removal of a Zn atom from the pristine ZnS structure, several new absorption peaks were observed. In Fig.\ref{fig4}(b), high absorption peaks at $0.8$ eV (infrared region (IR)), $3.2$ eV, and $3.6$ eV (both in the ultraviolet (UV) range) are shown. The higher intensity of these components indicates that the neutral p-doped ZnS could readily absorb photons with such energies. In addition, the lower intensity absorption peaks are of particular interest, especially the component at $1.8$ eV, corresponding to visible range absorption. This result can be addressed to the formation of a p-type ZnS, highlighting the role of zinc vacancies in enhancing the optical absorption. This improvement was experimentally observed through the green-like emission upon excitation with a $405$ nm laser source, as shown in Fig.\ref{fig2}(b). It is worth noting that before emission, the sample must absorb the incoming photons from the laser source. This step requires an absorption level close to $3.06$ eV. Thus, in view of the new absorption peak at $3.2$ eV in the $V_{Zn}^{0}$-ZnS, this experimental finding might be attributed to the formation of neutral zinc vacancies. On the other hand, in the sight of PL experiment, the neutral zinc defect does not match the results. Once PL  first implies the absorption of photons with a particular source energy, a state at $2.54$ eV in the absorption spectra is required in our experiments. However, changing the charge state of the system to $q=-1$, rather than new components observed in the neutral scenario, a component at $2.7$ eV is found as revealed in Fig.\ref{fig4}(c), in good agreement with the experimental prerequisite. The charged p-type ZnS also displays absorption peaks in a broad range, from IR ($0.7$ eV, $1.08$ eV), visible (VIS) ($1.8$ eV, $2.2$ eV, and $2.7$ eV), and UV interval ($3.6$ eV and $3.8$ eV). It should be noted in the dipole approximation we evaluate the imaginary part of the dielectric function computing direct inter-band transitions by a summation over several empty states, neglecting any many-body effects. Therefore, slight energy differences can occur, as well as a loss of information, as widely discussed \cite{Li2015,Webster2015}. However, all signatures from this model should be retained through an improved methodology, which allows us to attest the good agreement between the experiments and first-principle calculations.

\begin{figure}[h!]
    \centering
    \includegraphics[width=1.0\textwidth]{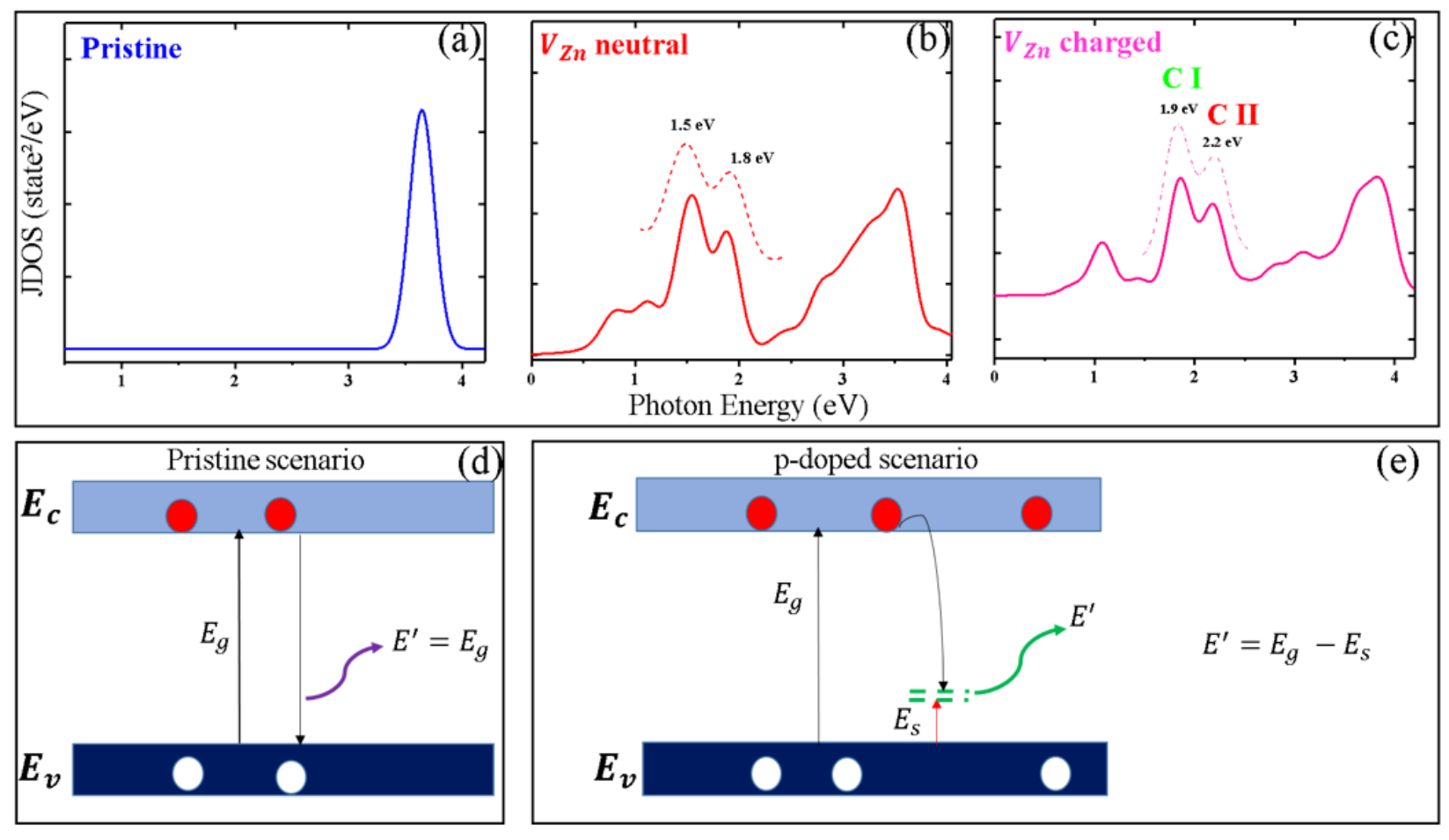}
    \caption{Joint Density of states (JDOS) of (a) pristine, (b) neutral $V_{Zn}$  and (c) charged $V_{Zn}$  ZnS. Pictorial diagram level illustrating the emissions in the (d) pristine and (e) p-doped scenarios.}
    \label{fig5}
\end{figure}

To address the new PL peaks, a suitable approach is to evaluate the joint density of states (JDOS), which represents the allowed inter-band transitions at specific energy levels. Essentially, it can be understood as an emission-like spectrum, enabling the correlation of defect-induced states with experimental photoluminescence features. Fig.\ref{fig5}(a)-(c) display the JDOS for both pristine and doped ZnS. Our analysis is confined to an energy range that includes IR, VIS, and UV-VIS regions. In the pristine system, there is only one emission, corresponding to a direct recombination process from the VBM to the CBM producing photons with $3.6$ eV. On the other hand, as we previously discussed, density of states (DOS) calculations show that when a p-type structure forms, new states slightly above the VBM emerge. These states can act as trapping sites, enabling recombination processes beyond the direct VBM-CBM transition. \cite{Zott1997}
To clarify the role of shallow levels in the emission spectra of ZnS, Fig. \ref{fig5}(b) to Fig.\ref{fig5}(c) show the JDOS of p-type ZnS, obtained by removing one zinc atom from the pristine supercell. In Fig.\ref{fig5}(b), the neutral configuration is evaluated, revealing new emission lines at $0.8$ eV, $1.5$ eV, and $1.8$ eV, in addition to the $3.6$ eV emission line. The high JDOS peaks at $1.5$ eV and $1.8$ eV suggest a higher probability of these emissions occurring. Therefore, we can infer that zinc vacancies not only enhance optical absorption but also improve optical emission. However, taking PL experiments as the reference, the neutral scenario alone cannot fully explain the two new emissions in the visible range, as discussed in the last section. Thus, we turn our attention to the charged system, depicted in Fig.\ref{fig5}(c). 

The features in this case are quite similar to those in the neutral structure, except for the energy of the emissions. The lowest probability of emission occurs at $1.2$ eV, while the highest JDOS components, apart from the $3.6$ eV peak, fall exactly in the visible range at $1.9$ eV (C I) and $2.2$ eV (C II). This finding aligns well with PL experiments, which revealed two new peaks at $2.05$ eV and $2.12$ eV. Therefore, our ab initio calculations confirm zinc vacancies broadening the ZnS emission lines and suggest these vacancies might be charged in the system. Indeed, as discussed in terms of the formation energy, both neutral and charged zinc vacancies are stable in zb-ZnS and may coexist. The pictorial model distinguishing the pristine and doped scenario is illustrated in Fig.\ref{fig5}(d) to Fig.\ref{fig5}(e). In the pristine scenario (Fig. \ref{fig5}(d), an electron is excited from the valence band to the conduction band. Sequentially, this free excited electron recombines directly into the valence band, emitting a photon with energy $E$, which is equal to the band gap of the system. In contrast, in p-type-like structures, the additional shallow levels in the forbidden zone act as trapping sites. In this case, the free electron that jumped from VBM to CBM recombines at the impurity level $E_{s}$, resulting in an emission whose energy $E'$ is the difference of the band gap and shallow level energy.

%% file: conclusion.tex
\section{Conclusion}
The electronic and optical response of zb- ZnS were explored by combining experimental and theoretical tools. The presence of zinc vacancies, as suggested by XPS and PL analysis, is the driving force for improving the electronic behavior and broadening the optical absorption range. In the defective scenario, the charging of the sample was mitigated (electronic response improvement) and new emission lines in the visible range were found (optical response enhancement). The first finding is explained by the formation of new levels inside the gap, due to the removal of zinc atoms, as supported by electronic structure calculation using the PBE-GGA methodology. The new lines revealed from PL experiments are explained based on the radiative recombination process: zinc vacancies introduce shallow levels that act as trapping sites, allowing the recombination processes to occur in addition to the direct VBM-CBM transition. Such experimental findings are supported by  absorption spectra and JDOS calculations through random phase approximation (RPA) employing the PBE-$\alpha$ methodology. Owing to the stable p-type configuration, a wide range of absorption (including IR, VIS and UV) is possible, and new emission lines of $V_{Zn}$-ZnS are observed.  Our results highlight the key role of zinc vacancies in zb-ZnS in tuning the optical and electronic properties of ZnS, making it a promising candidate for photodetector, photocatalysis, and light-emitting applications. Future studies could explore hybrid ZnS/organic interfaces in the presence of zinc vacancies to further understand the role of cation defects on the interaction with organic molecules like CO and $CO_{2}$.

%% file: FurtherParts/AuthorContribution.tex
\section*{CRediT authorship contribution statement}  
\textbf{P.R.A. de Oliveira}: Formal analysis, Conceptualization, Methodology, Data curation, Writing - original draft, Writing - review \& editing.  
\textbf{L. Lima}: Formal analysis, Writing - review \& editing.  
\textbf{G. Félix}: Formal analysis, Writing - review \& editing.  
\textbf{P. Venezuela}: Formal analysis, Methodology, Writing - review \& editing.  
\textbf{F. Stavale}: Formal analysis, Conceptualization, Methodology, Project administration, Writing - review \& editing

%% file: FurtherParts/AuthorDeclaration.tex
\section*{Conflicts of interest}
The authors have no conflicts to disclose

%% file: FurtherParts/Acknowledgments.tex
\section*{Acknowledgments}
The authors thank the Conselho Nacional de Desenvolvimento Científico e Tecnológico (CNPq), Brazil, and the Fundação de Amparo à Pesquisa do Estado do Rio de Janeiro (FAPERJ) for financial support. P.R.A. de Oliveira and P. Venezuela also acknowledge the Centro Nacional de Processamento de Alto Desempenho (CENAPAD-SP) for providing computational resources. Additionally, P.R.A. de Oliveira acknowledges E.V.C. Lopes and Shyam B. Patel for critically reviewing the manuscript.

%% file: FurtherParts/SupportingInformation.tex
\section*{Supporting Information}
The following information can be found in the supporting Information : (1) General Information regarding ZnS ; (2) Optical properties descriptions - JDOS and Kramer-Kronig equations ; (3) Laser-assisted VB-XPS spectra ; (4) Chemical potential limit and Formation Energy equation ; (5) Transition energy states of $V_{Zn}$ and $V_S$ ZnS - Main equation and short comments.

%% file: FurtherParts/DataAvailability.tex
\section*{Data availability}
All relevant data are included in the paper and the Supporting Information.